\documentclass[iop]{emulateapj}
\newcommand{\kms}{km~s$^{-1}$\ }
\usepackage{url}
\shorttitle{A unified dynamical scaling relation}
\shortauthors{Cortese et al.}

\begin{document}


\title{The SAMI Galaxy Survey:\\ Towards a unified dynamical scaling relation for galaxies of all types}

 
%\author{L. Cortese\altaffilmark{1}, et al.}
%\affil{$^{1}$Centre for Astrophysics \& Supercomputing, Swinburne University of Technology, Mail H29 - P.O. Box 218, Hawthorn, 3122 Victoria, Australia}

\author{L. Cortese\altaffilmark{1}, L.~M.~R. Fogarty\altaffilmark{2,3}, I.-T. Ho\altaffilmark{4}, K. Bekki\altaffilmark{5}, J. Bland-Hawthorn\altaffilmark{2,3}, 
M. Colless\altaffilmark{6}, W. Couch\altaffilmark{7}, S.~M. Croom\altaffilmark{2,3}, 
K. Glazebrook\altaffilmark{1}, J. Mould\altaffilmark{1}, N. Scott\altaffilmark{2,3}, R. Sharp\altaffilmark{6}, C. Tonini\altaffilmark{8}, 
J.~T. Allen\altaffilmark{2,3}, 
J. Bloom\altaffilmark{2,3},
J.~J. Bryant\altaffilmark{2,3,7}, 
M. Cluver\altaffilmark{9}, 
R.~L. Davies\altaffilmark{10},
M. Drinkwater\altaffilmark{11}, 
M. Goodwin\altaffilmark{7}, 
A. Green\altaffilmark{7}, 
L.~J. Kewley\altaffilmark{6}, 
I.~S. Kostantopoulos\altaffilmark{7,3}, 
J.~S. Lawrence\altaffilmark{7}, 
S. Mahajan\altaffilmark{11,12}, 
A.~M. Medling\altaffilmark{6},
M. Owers\altaffilmark{7}, 
S.~N. Richards\altaffilmark{2,7,3},
S.~M. Sweet\altaffilmark{6}, 
O.~I. Wong\altaffilmark{5}}

\affil{$^{1}$Centre for Astrophysics \& Supercomputing, Swinburne University of Technology, Hawthorn, 3122 Victoria, Australia\\
$^{2}$ Sydney Institute for Astronomy (SIfA), School of Physics, The University of Sydney, NSW 2006, Australia\\
$^{3}$ ARC Centre of Excellence for All-sky Astrophysics (CAASTRO)\\
$^{4}$ Institute for Astronomy, University of Hawaii, 2680 Woodlawn Drive, Honolulu, HI 96822, USA\\
$^{5}$ ICRAR, The University of Western Australia, Crawley WA 6009, Australia\\
$^{6}$ Research School of Astronomy \& Astrophysics, The Australian National University, Cotter Road, Weston Creek, ACT 2611, Australia\\
$^{7}$ Australian Astronomical Observatory, PO Box 915, North Ryde, NSW 1670, Australia\\
$^{8}$ School of Physics, The University of Melbourne, VIC 3010, Australia\\
$^{9}$ Astronomy Department, University of Cape Town, Private Bag X3, Rondebosch7701, South Africa\\
$^{10}$ Astrophysics, Department of Physics, University of Oxford, Denys Wilkinson Building, Keble Rd., Oxford OX1 3RH, UK\\
$^{11}$ School of Mathematics and Physics, University of Queensland, QLD 4072, Australia\\
$^{12}$Indian Institute of Science Education and Research Mohali-IISERM, Knowledge City, Sector 81, Manauli, PO 140306, India}
%


%\altaffiltext{1}{Visiting Astronomer, Cerro Tololo Inter-American Observatory.
%CTIO is operated by AURA, Inc.\ under contract to the National Science
%\altaffiltext{2}{Society of Fellows, Harvard University.}
%\altaffiltext{3}{present address: Center for Astrophysics,
%    60 Garden Street, Cambridge, MA 02138}
%\altaffiltext{4}{Visiting Programmer, Space Telescope Science Institute}
%\altaffiltext{5}{Patron, Alonso's Bar and Grill}


\begin{abstract}
We take advantage of the first data from the Sydney-AAO Multi-object Integral field (SAMI) Galaxy Survey 
to investigate the relation between the kinematics of gas and stars, and stellar mass 
in a comprehensive sample of nearby galaxies. 
We find that all 235 objects in our sample, regardless of their morphology, lie on a tight relation 
linking stellar mass ($M_{*}$) to internal velocity quantified by the $S_{0.5}$ parameter, 
which combines the contribution of both dispersion ($\sigma$) and rotational velocity ($V_{rot}$) 
to the dynamical support of a galaxy  ($S_{0.5}=\sqrt{0.5V_{rot}^{2}+\sigma^{2}}$). 
Our results are independent of the baryonic component from which $\sigma$ and $V_{rot}$ 
are estimated, as the $S_{0.5}$ of stars and gas agree remarkably well. 
This represents a significant improvement compared to the canonical $M_{*}$ vs. $V_{rot}$ and 
$M_{*}$ vs. $\sigma$ relations. Not only is no sample pruning necessary, but also 
stellar and gas kinematics can be used simultaneously, as the effect of asymmetric drift 
is taken into account once $V_{rot}$ and $\sigma$ are combined. 
%Intriguingly, we find evidence for a change in the slope of the $M_{*}$ vs. $S_{0.5}$ 
%relation for M$_{*}<$10$^{10}$ M$_{\odot}$, 
%as expected if $S_{0.5}$ linearly correlates with the total baryonic mass of galaxies. 
Our findings illustrate how the combination of dispersion and rotational velocities for both gas and stars 
can provide us with a single dynamical scaling relation valid for galaxies of all morphologies 
across at least the stellar mass range 8.5$<log(M_{*}/M_{\odot})<$11. 
Such relation appears to be more general and at least as tight as any other dynamical scaling relation, 
representing a unique tool for investigating the link between galaxy kinematics and baryonic content, and 
a less biased comparison with theoretical models. 
\end{abstract}





\section{Introduction}
It is well established that the internal velocity of disk \citep{tully77}
and spheroidal galaxies \citep{faber76} scales with their luminosity, stellar 
and baryonic mass \citep{mcgau00}. 
In addition to being important secondary distance indicators, 
the Tully-Fisher and Faber-Jackson relations provide strong 
constraints on galaxy formation and evolution (e.g., \citealp{baugh2006}).

Unfortunately, both relations 
hold only for accurately pre-selected classes of objects (e.g., inclined 
disks and bulge-dominated systems, respectively), and their scatters 
and slopes vary when wider ranges of morphologies are considered (e.g., \citealp{neinstein97,iodice2003,william10,catinella12,tonini14}). 
This limitation has hampered the comparison with theoretical models, 
as it is challenging to apply the same selection 
criteria used for observations to simulated data. 

Thus, recent works have started investigating the possibility of bringing galaxies 
of all morphologies onto the same dynamical scaling relation. 
\cite{kassin2007} showed that, once the contributions of rotation ($V_{rot}$) and dispersion ($\sigma$) of 
the H$\alpha$-emitting gas are combined into the $S_{0.5}$ parameter ($S_{0.5}=\sqrt{0.5V_{rot}^{2}+\sigma^{2}}$, 
\citealp{weiner06}), all star-forming galaxies (including merging systems) lie on a tight 
($\sim$0.1 dex scatter) stellar mass ($M_{*}$) vs. $S_{0.5}$ relation. 
Although it is still debated whether the combination of 
$V_{rot}$ and $\sigma$ is necessary to reduce the scatter in the Tully-Fisher relation of late-type galaxies 
(including disturbed systems, \citealp{miller11}), it is intriguing that the slope and intercept 
of the $M_{*}$ vs. $S_{0.5}$ relation found by \cite{kassin2007} 
is close to that of the Faber-Jackson relation, suggesting that a similar approach might hold also for quiescent systems. 

\cite{zaritsky2008} addressed this issue by using the $S_{0.5}$ parameter 
to show that ellipticals and disk galaxies lie on the same scaling relation. 
However, contrary to \cite{kassin2007} who directly combined 
$\sigma$ and $V_{rot}$, they simply used rotational velocities for disks 
and integrated dispersion velocities for bulges. As in massive systems  
both rotation and dispersion contribute significantly to the dynamical support \citep{courteau2007,emsellem11}, 
these assumptions cannot be generalised to the entire population of galaxies. 

\cite{catinella12} were recently able to bring all massive ($M_{*}>10^{10}$ M$_{\odot}$) 
galaxies on a tight relation by using the galaxy's concentration index to correct 
the stellar velocity dispersion of disk-dominated systems. This empirical approach 
is motivated by the observed dependence of the $V_{rot}/\sigma$ ratio on morphology \citep{courteau2007}, 
suggesting that the $S_{0.5}$ parameter may indeed be applied to all types of galaxies. 

In this Letter, we combine gas and stellar kinematics for 235 galaxies observed as part of 
the Sydney-AAO Multi-object Integral field (SAMI, \citealp{croom12}) Galaxy Survey \citep{bryant14} 
to show that {\it all} galaxies lie on the same $M_{*}$ vs. $S_{0.5}$ relation. 
The major advantage of our approach lies 
%on the simultaneous combination of resolved dispersion and 
%rotation velocities maps of both gas and stars for all galaxies in our sample. 
in the measurement of dispersion and rotational velocities, from both stellar and gas components, from spatially resolved maps.

\section{The SAMI Galaxy Survey}

%% In a manner similar to \objectname authors can provide links to dataset
%% hosted at participating data centers via the \dataset{} command.  The
%% second curly bracket argument is printed in the text while the first
%% parentheses argument serves as the valid data set identifier.  Large
%% lists of data set are best provided in a table (see Table 3 for an example).
%% Valid data set identifiers should be obtained from the data center that
%% is currently hosting the data.
%%
%% Note that AASTeX interprets everything between the curly braces in the 
%% macro as regular text, so any special characters, e.g. "#" or "_," must be 
%% preceded by a backslash. Otherwise, you will get a LaTeX error when you 
%% compile your manuscript.  Special characters do not 
%% need to be escaped in the optional, square-bracket argument.

The SAMI Galaxy Survey is targeting $\sim$3400 galaxies in the redshift 
range 0.004$<z<$0.095 with the SAMI integral field unit, installed at the Anglo-Australian Telescope. 
Details on the target selection can be found in \cite{bryant14}. 
%The main goal 
%of this survey is to provide a complete census of the resolved optical 
%properties of nearby galaxies (e.g., age, metallicities, kinematics) 
%across a wide range of environments \citep{bryant14}.

SAMI takes advantage of photonic imaging bundles (`hexabundles', \citealp{bland11,hexa14}) 
to simultaneously observe 12 galaxies across a 1 degree field of view.
Each hexabundle is composed of 61 optical fibers, each with a diameter of $\sim$1.6\arcsec, 
covering a total circular field of view of $\sim$14.7\arcsec\ in diameter. 
SAMI fibers are fed into the AAOmega dual-beam spectrograph, providing 
a coverage of the 3700-5700 \AA\ and 6250-7350 \AA\ wavelength ranges at resolutions 
R$\sim$1750 and R$\sim$4500, respectively. 
These correspond to a full-width at half-maximum (FWHM) of $\sim$170 \kms in the blue, 
and $\sim$65 \kms in the red.

\subsection{Observations and data reduction} 
We focus on the first 304 galaxies observed by SAMI in the footprint of the Galaxy And Mass Assembly Survey (GAMA, \citealp{gama}) 
for the wealth of multiwavelength data available (see \S~\ref{data}). 

Observations were carried out on March 5-17 and April 12-16, 2013. The typical observing strategy 
consists of seven dithered observations totalling 3.5 hours to achieve near-uniform 
spatial coverage across each hexabundle. 
%The typical seeing during the observations was $\sim$2.5\arcsec (see also \citealp{allen14}), 
%corresponding to $\sim$2.1 kpc at the average redshift of our sample.
The AAOmega data reduction pipeline \textsc{2dfdr} was used to perform all the 
standard data reduction steps. Flux calibration was done taking 
advantage of a spectro-photometric standard star observed during the same night, 
while correction for telluric absorption was made using simultaneous observations 
of a secondary standard star (included in the same SAMI plate of the target).    
The row-stacked spectra of each exposure generated by \textsc{2dfdr} were then combined, reconstructed 
into an image and resampled on a Cartesian grid of 0.5\arcsec$\times$0.5\arcsec spaxel size 
(see \citealp{sharp14} and \citealp{allen14} for more details). 
%on the flux calibration and resampling procedure).

\subsection{Stellar and gas kinematics}
To obtain homogenous global rotation and dispersion velocities for both gas and stars within one effective radius ($r_{e}$), 
we select the 250 galaxies in our sample with an $r$-band effective diameter (see \S~\ref{data}) smaller than the size of a SAMI hexabundle 
(14.7\arcsec), and greater than the typical spatial resolution of our observations (2.5\arcsec, i.e., $\sim$2.1 kpc 
at the average redshift of our sample, see also \citealp{allen14}). 
The average diameter of our sample is $\sim$8\arcsec.
 
\begin{figure*}
\epsscale{1.14}
\plotone{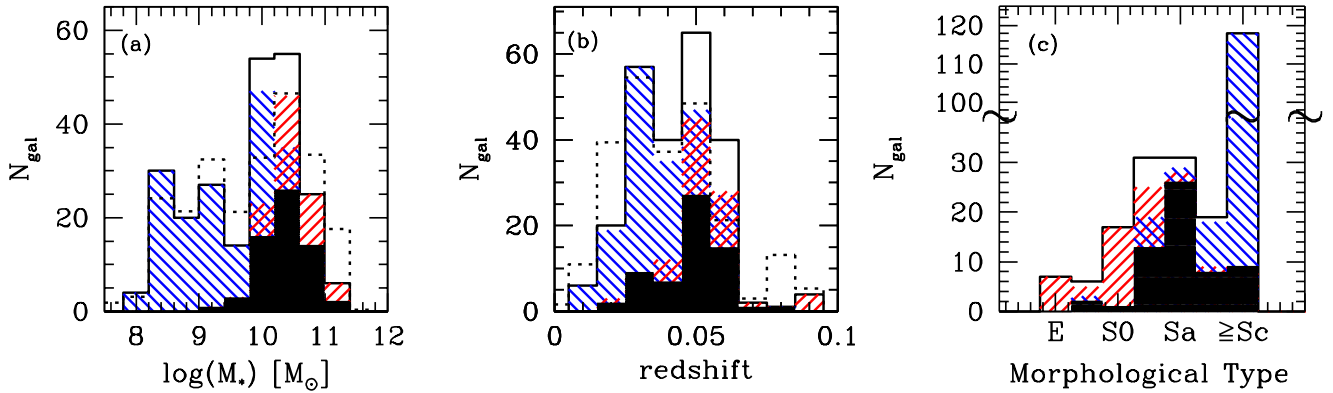}
\caption{\label{fig1} The stellar mass (a), redshift (b) and morphological type (c) distributions for our sample (solid histograms). 
Blue, red and filled histograms indicate galaxies with gas, stellar, and both gas and stellar kinematics, respectively. 
The dotted histogram in (a) and (b) shows the distribution (scaled to the size of our sample) of the SAMI primary sample (see Bryant et al. 2014b).}
\end{figure*}
Stellar line-of-sight velocity and intrinsic dispersion maps were then obtained using 
the penalised pixel-fitting routine \textsc{ppxf}, developed by \cite{cappellari2004}, 
following the same pixel-by-pixel technique described in \cite{fogarty14}.

%We use the 985 stellar template 
%spectra from the MILES stellar library \citep{miles}, and fit only pixels in the data cube 
%with signal-to-noise per spectral pixel greater than 5 following \cite{fogarty14}.

Gas velocity maps were created from the reduced data cubes using the 
new \textsc{lzifu} IDL fitting routine (Ho et al. in prep.; see also \citealp{ho14}). 
After subtracting the stellar continuum with \textsc{ppxf}, \textsc{lzifu} models 
the emission lines in each spaxel as Gaussians and performs a non-linear least-square fit using the Levenberg-Marquardt method. 
We fit up to 11 strong optical emission lines ([O{\sc ii}]$\lambda\lambda$3726,29, H$\beta$, [O{\sc iii}]$\lambda\lambda$4959,5007, 
[O{\sc i}]$\lambda$6300, [N{\sc ii}]$\lambda\lambda$6548,83, H$\alpha$, and [S{\sc ii}]$\lambda\lambda$6716,31) simultaneously, constraining 
all the lines to share the same rotation velocity and dispersion. 
Each line is modelled as a single-component Gaussian (including the effect of instrumental resolution; e.g., \citealp{weiner06}), 
and we use the reconstructed kinematic maps to measure gas rotation and intrinsic velocity dispersion.

%We limit our analysis to those galaxies with `high-quality' kinematics for 
%$>$80\% of the spaxels included within an ellipse of major semi-axis $r_{e}$ and ellipticity and position angle 
%determined from optical $r$-band Sloan Digital Sky Survey (SDSS, \citealp{york2000}) images (see \S~\ref{data}).
%Spaxels are discarded if the fit failed or if the error associated to the velocities 
%is greater than 20 \kms and 50 \kms for gas and stars, respectively. This conservative cut 
%roughly corresponds to one third of FWHM in SAMI cubes.

We then select our final sample as follows. 
First, spaxels are discarded if the fit failed or if the error on the velocities 
is greater than 20 \kms and 50 \kms for gas and stars, respectively. This conservative cut 
roughly corresponds to one third of spectral FWHM in SAMI cubes.
Second, we estimate the fraction of `good' spaxels ($f$) left within an ellipse of semi-major axis $r_{e}$ and ellipticity and position angle 
determined from optical $r$-band Sloan Digital Sky Survey (SDSS, \citealp{york2000}) images (see \S~\ref{data}), 
and reject those galaxies with $f<$80\%. 
This selection guarantees that we are properly tracing the galaxy kinematics up to $r_{e}$, and 
it leaves us with 235 individual galaxies: 193 with 
gas kinematics, 105 with stellar kinematics and 62 with both (see Fig.~\ref{fig1}). 
Although our analysis takes advantage of 
less than 10\% of the final SAMI Galaxy Survey, 
the sample size is already comparable to the largest IFU surveys 
of nearby galaxies to date \citep{cappellari11,califa}. 
The properties of our final sample are summarised in Fig. \ref{fig1}.

Stellar and gas velocity widths ($W$) are obtained from the velocity histogram created 
by combining all the `good' spaxels within $r_{e}$. Following the 
standard technique used for H$\alpha$ rotation curves, we define $W$ as the difference between the 90th and 
10th percentile points of the velocity histogram ($W=V_{90}-V_{10}$, \citealp{cati05}). 
We adopt the velocity histogram technique because this is the simplest method 
to determine velocity widths, making our results easily comparable to other studies, including 
long-slit spectroscopy.    
Rotational velocities are then computed as 
\begin{equation}
V_{rot} = \frac{W}{2(1+z)sin(i)}
\end{equation}    
where $i$ is the galaxy inclination and $z$ is the redshift. 
Inclinations are determined from the $r$-band minor-to-major axis ratio ($b/a$) as:
\begin{equation}
cos(i)=\sqrt{\frac{(b/a)^{2}-q_{0}^{2}}{1-q_{0}^{2}}}
\end{equation}    
where $q_{0}$ is the intrinsic axial ratio of an edge-on galaxy. 
Following \cite{catinella12}, we adopt $q_{0}$=0.2 for all galaxies and set 
to inclination of 90 degrees if $b/a<$0.2. Our conclusions are unchanged if 
we vary $q_{0}$ with morphology. The average $b/a$ of our sample is $\sim$0.5.

Stellar and gas velocity dispersions are defined as the linear average of the 
velocity dispersion measured in each `good' spaxel (without any correction for inclination). 
We preferred linear to luminosity-weighted 
averages to be consistent with our velocity width measurements (which are not luminosity-weighted) 
and because these are less affected by beam smearing \citep{davies2011}. Our 
conclusions, however, are unchanged if we use luminosity-weighted quantities. 
Excluding the effect of inclination, we assume an uncertainty of 0.1 dex  
in the estimate of both $V_{rot}$ and $\sigma$.

\begin{figure*}
\epsscale{1.14}
\plotone{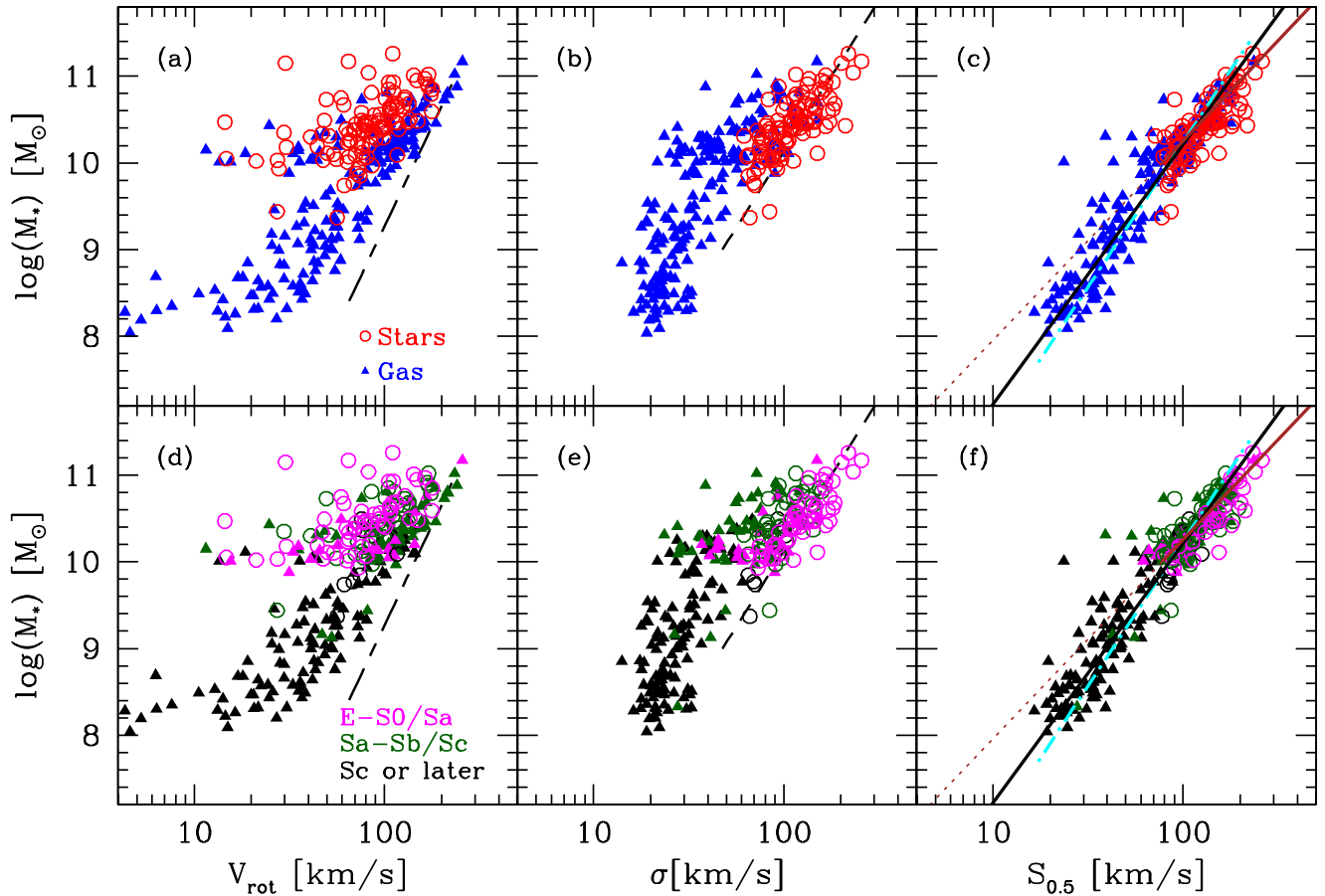}
\caption{\label{fig2} The $M_{*}$ vs. $V_{rot}$ (left), $\sigma$ (center) and $S_{0.5}$ (right) relations for our sample. 
Circles and triangles indicate stellar and gas kinematics, respectively. In the bottom row, symbols are colour-coded according 
to morphological type: E-S0/Sa (magenta), Sa-Sb/Sc (dark green), Sc or later types (black). The stellar mass Tully-Fisher (\citealp{bell2001}, long-short dashed line) 
and Faber-Jackson (\citealp{gallazzi06}, dashed line) relations for nearby galaxies are shown for comparison. 
In the right panels, the black solid line indicates the best inverse linear fit for the whole sample. The brown solid line is 
the best fit for $M_{*}>$10$^{10}$ M$_{\odot}$ only, with its extrapolation to lower masses shown as dotted line. 
The dashed-dotted line is the $M_{*}$ vs. $S_{0.5}$ relation of \cite{kassin2007}.}
\end{figure*}

We combine dispersion and rotation through the $S_{K}$ parameter: 
\begin{equation}
S_{K} = \sqrt{K V_{rot}^{2} + \sigma^{2}}
\end{equation}    
As discussed in \cite{weiner06} and \cite{kassin2007}, this quantity includes the dynamical support 
from both ordered and disordered motions and, thus, should be a better proxy for 
the global velocity of the galactic halo. 
Moreover, it is almost unaffected by beam smearing, as the artificial increase of 
$\sigma$ and decrease of $V_{rot}$ compensate each other once they 
are combined into $S_{K}$ \citep{covington10}. 

%As the typical seeing of these first SAMI observations is $\sim$2.5\arcsec (corresponding to $\sim$2.1 kpc at the average 
%redshift of our sample), beam smearing may indeed affect our estimates of $V_{rot}$ and $\sigma$, 
%urther supporting the use of the $S_{0.5}$ parameter to trace the dynamical mass of galaxies. 
%of both rotation and dispersion velocities notoriously affect As shown by \cite{convigton10}, the addition power of the $S_{0.5}$ is that it is less affected 
%by  
Although the value of $K$ varies with the properties of the system, in this paper we follow 
the simple approach of \cite{kassin2007} and \cite{zaritsky2008}, and fix $K$=0.5. 
Our  conclusions do not change for 0.3$<K<$1.

\subsection{Ancillary data}
\label{data}
The SAMI data are combined with multiwavelength observations obtained as part of the GAMA survey.
Briefly, $r$-band effective radii, position angles and ellipticities are taken from the one-component 
Sersic fits presented in \cite{kelvin12}\footnote{We re-computed the ellipticity and position angle for seven galaxies with bright bars or other issues (GAMA 250277, 
279818, 296685, 383259, 419632, 536625, 618152) as the published values do not match the orientation of the velocity field.}. 
Stellar masses ($M_{*}$) are estimated from $g-i$ colours and $i$-band magnitudes following \citet[see also \citealp{bryant14}]{taylor11}.
%, and stellar mass surface densities ($\mu_{*}$) are defined as $M_{*}/(2\pi R_{50,z}^{2})$, where $R_{50,z}$ is 
%the SDSS Petrosian radius in $z$-band. Finally, GALEX \citep{martin05} near-ultraviolet (NUV) magnitudes are taken 
%from the GAMA database\footnote{\url{http://www.gama-survey.org/dr2/}}.  

%Visual morphological classification has been performed on the SDSS RGB colour images. Nine of us independently 
%classified each galaxy following the scheme used by \cite{kelvin14}. Galaxies are first divided into late- and early-types 
%according to their shape, presence of spiral arms and/or signs of star formation. Then, early-types with just a bulge are classified as ellipticals (E), 
%whereas early-types with disks are S0s. Similarly, late-type galaxies with a bulge component are Sa-Sb, whereas bulge-less late-type galaxies are Sc or later.
%All votes are then combined and, for each galaxy, the type with at least 66\% of the votes is chosen. 

Visual morphological classification has been performed on the SDSS colour images, following the scheme used by \cite{kelvin14}. 
Galaxies are first divided into late- and early-types according to their shape, presence of spiral arms and/or signs of star formation. 
Then, early-types with just a bulge are classified as ellipticals (E), whereas those with disks as S0s. Similarly, late-type galaxies 
with a bulge component are Sa-Sb, whereas bulge-less late-type galaxies are Sc or later.
%All votes are then combined and, for each galaxy, the type with at least 66\% of the votes is chosen. 

\section{Dynamical scaling relations}
Fig.~\ref{fig2} shows $M_{*}$ vs. $V_{rot}$ (left panel), $\sigma$ (middle) and $S_{0.5}$ (right) for all 235 galaxies in our sample. 
Circles and triangles indicate galaxies with kinematical parameters from stellar and 
gas components, respectively. Thus, the 62 galaxies for which both gas and stellar kinematics are available appear 
twice in each plot.  In the bottom row, galaxies are colour-coded according to their morphology. 

\subsection{The stellar mass Tully-Fisher relation}
Our $M_{*}$ vs. $V_{rot}$ relation has a larger scatter ($\sim$0.26 dex in $V_{rot}$ from the inverse fit)\footnote{All scatters in 
this paper are estimated from the inverse linear fit along the x-axis: i.e., we consider $M_{*}$ as independent variable.} than 
classical Tully-Fisher relation ($\sim$0.08 dex). This is not surprising as our sample includes early-types and face-on systems that would normally be excluded 
from Tully-Fisher studies (e.g., \citealp{catinella12}). Indeed, a significant fraction of the scatter is due to spirals with bulges, and early types (see Fig.~\ref{fig2}d). 

\begin{figure*}
\epsscale{1.14}
\plotone{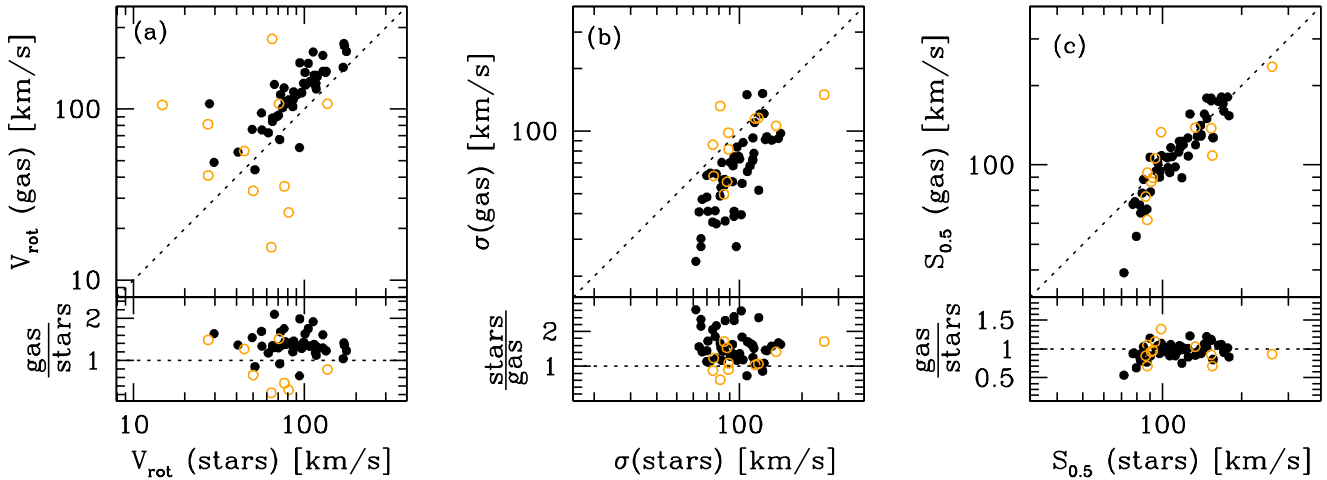}
\caption{\label{fig3} Comparison between rotation ($V_{rot}$, left), dispersion velocities ($\sigma$, center) and $S_{0.5}$ (right) of gas and stars, for the 
62 galaxies in our sample with both measurements available. Empty circles highlight galaxies where gas and stars 
have a misaligned rotation axis. In each panel, the dotted line shows the 1-to-1 relation.}
\end{figure*}

The $M_{*}$ vs. $V_{rot}$  relation for the stars (circles) is significantly 
offset from the one of the gas (triangles): i.e., at fixed $M_{*}$, stars rotate slower than gas. 
This is clearer in Fig.~\ref{fig3}a, where we compare $V_{rot}$ 
of gas and stars for the 62 galaxies with both measurements available. 
Once galaxies with clear misalignments between gas and stellar rotation axis are excluded (empty circles), 
we find that $V_{rot}$(gas) is, on average, $\sim$0.14 dex (with standard deviation $SD \sim$0.11 dex) higher than $V_{rot}$(stars). 
This is a consequence of asymmetric drift: while gas and stars experience the same galactic 
potential, a larger part of the stellar dynamical support comes from dispersion.
The average ratio $V_{rot}$(stars)/$V_{rot}$(gas) is 0.75 ($SD\sim$0.20), roughly $\sim$20\% lower than the value obtained by \cite{martinsson2013} by comparing 
the maximum rotational velocities of pure disk galaxies ($V_{rot}$(stars)/$V_{rot}$(gas)$\sim$0.89). 
This is likely due to the fact that our sample is mainly composed by early-type spirals (see Fig.~\ref{fig1}c) and that we probe 
only the central parts of galaxies, where asymmetric drift is more prominent.

For comparison, in Fig.~\ref{fig2}a,d, we show the local stellar mass Tully-Fisher relation 
\citep{bell2001}\footnote{Stellar masses have been converted to a Chabrier Initial Mass Function following \cite{gallazzi08}.}.
Our relation is flatter, showing a good match only at high $M_{*}$. 
This is because our rotational velocities are measured within $r_{e}$. 
As the rotation curves of giant galaxies rise more quickly than in dwarfs \citep{catinella2006}, 
our $V_{rot}$ are close to the maximum rotational velocity only in massive systems (see 
also Fig.~2 in \citealp{yegorova07}).  \\

%  value is larger than what obtained by comparing the maximum rotational velocities  
%The difference between gas and stars decreases to 15\% (consistent with ???) if we compare the maximum velocity widths 
%measured within $r_{e}$, confirming that asymmetric drift is more prominent in the inner regions.

\subsection{The stellar mass Faber-Jackson relation}
As for the $M_{*}$ vs. $V_{rot}$ relation, the scatter of our $M_{*}$ vs. $\sigma$ relation is larger ($\sim$0.16 dex) 
than the one typically obtained for early-type galaxies only ($\sim$0.07 dex, \citealp{gallazzi06}). As \cite{catinella12}, we find 
that the offset from the $M_{*}$ vs. $\sigma$(stars) relation for early-type galaxies (\citealp{gallazzi06}; dashed line) 
correlates with the concentration index.  
%[$\Delta\sigma{\rm(stars)} = (-0.41\pm0.07) + (0.14\pm0.03)\times(R_{90}/R_{50})$]. 
This confirms that, at fixed $M_{*}$, disks are more rotationally supported than bulge-dominated systems.  

The scaling relations of stars and gas are offset, with $\sigma$(gas) on average 0.19 dex ($SD\sim$0.13 dex) lower than $\sigma$(stars) 
(Fig.~\ref{fig3}b; see also \citealp{ho2009}).  In addition, the $M_{*}$ vs. $\sigma$ 
relation for the gas is not linear, as galaxies with M$_{*}<$10$^{10}$ M$_{\odot}$ have roughly the same velocity dispersion ($\sim$30 km~s$^{-1}$), i.e., 
the typical value observed in pure disk galaxies \citep{epinat08}. 

%Although this is the typical value observed in pure disk galaxies \citep{epinat08}, it is possible that 
%part of this sharp transition is an artefact due to the fact that we have reached the nominal SAMI 
%spectral velocity resolution in the red ($\sim$27 km~s$^{-1}$).   

\subsection{The $M_{*}$ vs. $S_{0.5}$ relation}
The large scatter and the difference between stars and gas observed for the $M_{*}$ vs. $\sigma$ and 
$M_{*}$ vs. $V_{rot}$ relations disappear when $V_{rot}$ and $\sigma$ are combined in the $S_{0.5}$ 
parameter (see Fig.~\ref{fig2}c,f). 
All morphological types follow the same scaling relation with just a few 
outliers. An inverse linear fit (assuming $S_{0.5}$ as dependent variable) gives 
a scatter of $\sim$0.1 dex (solid line in Fig.~\ref{fig2}c,f)\footnote{Note that galaxies with both $S_{0.5}$(gas) and $S_{0.5}$(stars) do not contribute 
twice to the fit, as for these we use the average between the two values.}. 
The slope and intercept of the linear relation (0.33$\pm$0.01, $-$1.41$\pm$0.08) are similar to what is found 
by \cite{kassin2007} for star-forming galaxies at $z\sim$0.1 (dotted-dashed line). 
This is interesting, as they used maximum rotational velocities, instead of 
velocities within $r_{e}$. 

The remarkable agreement between the $S_{0.5}$ for gas and stars is shown in Fig.~\ref{fig3}c: 
the average logarithmic difference (gas-stars) is just $\sim-$0.02 dex ($SD\sim$ 0.07dex), even including disturbed galaxies.
%\footnote{The only clear outlier (GAMA534654), which is also the strongest outlier of the $M_{*}$ vs. $S_{0.5}$  relation, is an almost perfectly face-on galaxy for which the inclination correction is highly %uncertain.}. 
This is expected if both quantities trace the potential of the galaxy, and justifies their 
combination on the same scaling relation. 
The agreement between gas and stars is little affected by the value 
of $K$ used to combine $V_{rot}$ and $\sigma$. Indeed, the average logarithmic difference varies 
between $-$0.06 and +0.03 dex for 0.3$<K<$1, and $SD$ stays roughly the same.
Fig.~\ref{fig3} confirms that the reduced scatter in the $M_{*}$ vs. $S_{0.5}$ relation is 
simply due to the fact that the combination of $V_{rot}$ and $\sigma$ provides a better 
proxy for the kinetic energy of a galaxy, which correlates with its mass.

%%simply a `numerical artefact' due to the fact that we are linearly combining two quantities in a log-log plot. 
%%Indeed, gas and stars are brought on the same scaling relation 
%%only once the effect of asymmetric drift is properly taken into account.  

%Gas and stars are brought on the same relation only by taking properly into account 
%It is well known that the scatter of a log-log plot can be reduced by simply linearly adding a constant 
%to one of the variables. Thus, it is natural to wonder whether the reduced scatter in the $M_{*}$ vs. $S_{0.5}$ is just a `numerical artefact'.  
%Fig.~\ref{fig2} clearly rules out the possibility that the reduced scatter in the $M_{*}$ vs. $S_{0.5}$ relation is simply a `numerical artefact' due to 
%the fact that we are linearly combining two quantities in a log-log plot. 

\begin{figure}
\epsscale{1.0}
\plotone{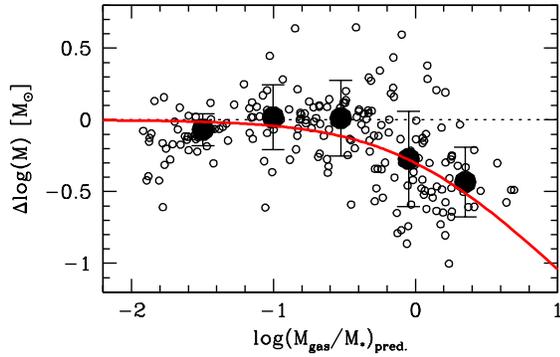}
\caption{\label{fig4} Residual along the y-axis ($M_{*}$) of the best fitting $M_{*}$ vs. $S_{0.5}$ relation for massive galaxies ($M_{*}>10^{10}$ M$_{\odot}$) 
as a function of `predicted' gas fraction (see text for details). Filled symbols show averages in bins of gas fraction. 
The red line is the trend expected if $S_{0.5}$ linearly correlates with total baryon mass.}
\end{figure}

Finally, we note that the $M_{*}$ vs. $S_{0.5}$ relation may become steeper 
for M$_{*}<$10$^{10}$ M$_{\odot}$. If we only fit massive galaxies (solid brown line in Fig.~\ref{fig2}c,f, with its extrapolation indicated as dotted line),  
low-mass systems appear systematically below the relation. 
This non linearity likely reflects the one observed in the $M_{*}$ vs. $\sigma$ relation, 
but it is tempting to speculate whether this is also somehow related to 
the similar feature observed in the stellar mass Tully-Fisher relation \citep{mcgau00}, 
which disappears in its baryonic version, once the mass of cold gas is taken into account. 
Unfortunately, the absence of cold gas measurements 
makes it impossible to compute a baryonic $S_{0.5}$ relation. 
Thus, we conclude by simply noting that, if we use the ultraviolet and optical properties of 
our systems to predict their total gas content\footnote{Atomic hydrogen masses 
are estimated following \cite{cortese11}. Total gas fractions are then obtained assuming a molecular-to-atomic gas ratio of 0.3 \citep{boselli14} and a helium
contribution of 30\%.}, 
the residuals from the linear fit vary with gas fraction (Fig.~\ref{fig4}) roughly as expected if the $S_{0.5}$ correlates 
linearly with total {\it baryonic} mass (red line in Fig.~\ref{fig4}). However, given all the assumptions and uncertainties, 
the idea that a linear baryonic $S_{0.5}$ relation might be the physical relation linking galaxies of all types remains 
for now a speculation.
 
 %Unfortunately, as the error in the predicted gas fractions is at least a factor of 3 larger than the scatter in the $M_{*}$ vs. $S_{0.5}$, 
%we cannot investigate if the inclusion of gas masses further reduces the dispersion in this relation.    

%
% This non linearity might simply reflect the one observed in the $M_{*}$ vs. $\sigma$ relation, 
% but a similar feature is also observed in the stellar mass Tully-Fisher 
% relation of pure disk galaxies \citep{mcgau00}, and disappears in its baryonic version, which accounts 
% for the mass of cold gas in galaxies. Unfortunately, the absence of cold gas measurements 
% makes it impossible to compute a baryonic $S_{0.5}$ relation. 
% Thus, we conclude by simply noting that, if we use the ultraviolet and optical properties of 
% our systems to predict their total gas content\footnote{Atomic hydrogen masses 
% are estimated following \cite{cortese11}. Total gas fractions are then obtained assuming a molecular-to-atomic gas ratio of 0.3 \citep{boselli14} 
% and a helium contribution of 30\%}, 
% the residuals from the linear fit correlate with gas fraction (Fig.~\ref{fig4}) roughly as expected if the $S_{0.5}$ correlates 
% linearly with total {\it baryonic} mass (red line in Fig.~\ref{fig4}). However, given all the assumptions and uncertainties, 
% the idea that a linear baryonic $S_{0.5}$ relation might be the physical relation linking galaxies of all types, {\bf remains 
% for the time being just a mere speculation.}
%

\section{Conclusions}
We take advantage of the first large statistical 
sample observed by the SAMI Galaxy Survey to show that all galaxies, regardless of 
their morphology, follow a tight ($\sim$0.1 dex) dynamical scaling relation once 
their dynamical support is expressed by combining the contributions of both 
rotational and dispersion velocities. 
%the $S_{0.5}$ parameter. 
We highlight that, while the stellar and gas components 
show systematic differences in their rotational and dispersion 
velocities, their $S_{0.5}$ agree remarkably well. 
This justifies the simultaneous use of both 
gas and stellar kinematical indicators, allowing us to bring 
both star-forming and quiescent systems on the same physical relation.

Our analysis improves on \cite{kassin2007} by showing that 
quiescent objects follow the same $M_{*}$ vs. $S_{0.5}$ relation 
as star-forming systems. To our knowledge, this is the first time 
that gas and stellar $V_{rot}$ and $\sigma$ 
for galaxies of all morphologies are combined on one 
dynamical scaling relation. This is a significant step forward compared 
to \cite{zaritsky2008}, confirming that the $S_{0.5}$ parameter 
can be applied to all types of galaxies. 
%%In addition, we present evidence for a possible change in the slope of 
%%the $M_{*}$ vs. $S_{0.5}$ relation at low stellar masses.
%, 
%as previously observed for the stellar mass Tully-Fisher relation, 
%suggesting that $S_{0.5}$ may linearly scale with the total baryonic 
%mass of galaxies.
 
%We show that the offset from a linear relation (in log-log space) correlates 
%with the predicted cold gas mass fraction, 
%as expected if $S_{0.5}$ scales linearly 
%with the total baryonic mass of galaxies. 

The $S_{0.5}$ parameter works remarkably well not only because it 
combines the contributions of both $V_{rot}$ and $\sigma$ to the dynamical  
support of galaxies, but also because it is influenced only weakly by instrumental 
effects \citep{covington10}. 
%While here we use $K$=0.5, our main 
%conclusions remain unchanged for 0.3$<K<$1. 

It is important to remember that, as already known 
for the Tully-Fisher and Faber-Jackson relations, the 
slope and scatter of the $M_{*}$ vs. $S_{0.5}$ relation likely depend on 
the technique used to estimate $V_{rot}$ and $\sigma$, as well as on 
the radius at which both quantities are measured. 
We plan to investigate this further with the full 
SAMI sample, as determining the best combination of $V_{rot}$ and $\sigma$ 
should reveal important information on the kinematical structure of galaxies.

In the meantime, the absence of any pre-selection in the sample 
not only makes the $S_{0.5}$ parameter extremely promising for characterising the dynamical properties of galaxies, but also 
might allow a more rigorous comparison with theoretical models.

\acknowledgments
We thank the referee for a constructive report. 
LC acknowledges support from the Australian Research Council (DP130100664).
ISK is the recipient of a John Stocker Postdoctoral Fellowship from 
the Science and Industry Endowment Fund (Australia).
The SAMI Galaxy Survey is based on observations made at the Anglo-Australian Telescope. 
The Sydney-AAO Multi-object Integral field spectrograph (SAMI) was developed jointly by the 
University of Sydney and the Australian Astronomical Observatory. The SAMI input catalogue is 
based on data taken from the Sloan Digital Sky Survey, the GAMA Survey and the VST ATLAS Survey. 
The SAMI Galaxy Survey is funded by the Australian Research Council Centre of Excellence for 
All-sky Astrophysics (CAASTRO), through project number CE110001020, and other participating institutions. 
%The SAMI Galaxy Survey website is http://sami-survey.org/


%% in parentheses, after the keyword, but they will not be verified.

{\it Facilities:} \facility{AAO/SAMI}.

\end{document}